\documentclass[11pt]{article}
\usepackage[colorlinks=true]{hyperref}
\hypersetup{colorlinks=true,linkcolor=teal,citecolor=blue,urlcolor=blue}
\usepackage{graphicx}
\usepackage{amsmath}
\usepackage{amssymb}
\usepackage{latexsym}
\usepackage{color}
\usepackage{subfigure}
\usepackage{pst-all}
\usepackage{pstricks,pst-node,pst-text,pst-3d}
\usepackage{ifpdf}
\usepackage{multimedia}
\usepackage{lmodern}
\usepackage{cite}
\usepackage{dsfont}
\usepackage{bbm}
\usepackage{lipsum}
\usepackage{url}
\usepackage{textcomp}
\usepackage{bbold}
\usepackage{epstopdf}
\usepackage{epsfig}
\usepackage{mathtools}
\usepackage{nccmath}
 \usepackage[normalem]{ulem}
\usepackage{ifpdf}
\ifpdf
\else
  \usepackage{pstricks}
\fi
  \hypersetup{pdfstartview=XYZ}
\newcommand{\bra}{\begin{array}}
\newcommand{\era}{\end{array}}
\newcommand{\beq}{\begin{equation}}
\newcommand{\eeq}{\end{equation}}
\newcommand{\beqar}{\begin{eqnarray}}
\newcommand{\eeqar}{\end{eqnarray}}

\def\BC{\bb C}
\def\_\BC{\bbi C}

\def\( {\left(}
\def\) {\right)}
\def\[ {\left[}
\def\] {\right]}
\def\no2 {{\textstyle{n\over 2}}}

\begin{document}
\thispagestyle{empty}
\begin{center}

\vspace{1.8cm}
 \renewcommand{\thefootnote}{\fnsymbol{footnote}}
 
 {\Large {\bf Generalized uncertainty relation between thermodynamic variables in quantum thermodynamics}}\\

\vspace{1.5cm} {\bf Z. Abuali}$^{a}${\footnote {
email: {\sf Z.Abuali@uok.ac.ir}}}, {\bf F. H. Kamin}$^{a}${\footnote { email: {\sf
f.hatami@uok.ac.ir}}}, {\bf R. J. S. Afonso}$^{b}${\footnote { email: {\sf
ricardo.afonso@ifsc.usp.br}}}, {\bf D. O. Soares-Pinto}$^{b}${\footnote { email: {\sf
dosp@ifsc.usp.br}}} and {\bf  S. Salimi}$^{a}${\footnote { email: {\sf
ShSalimi@uok.ac.ir}}}\\
\vspace{0.5cm}

$^{a}$ {\it  Department of Physics, University of Kurdistan, P.O.Box 66177-15175, Sanandaj, Iran }\\ [0.3em]
$^{b}$ {\it  Instituto de F\'isica de S\~ao Carlos, Universidade de S\~ao Paulo, CP 369, 13560-970, S\~ao Carlos, S\~ao Paulo, Brazil}\\ [0.3em]

\end{center}
\baselineskip=18pt
\medskip
\vspace{3cm}
\begin{abstract}
{Macroscopic thermodynamics, via the weak coupling approximation, assumes that the equilibrium properties of a system are not affected by interactions with its environment. However, this assumption may not hold for quantum systems, where the strength of interaction between the system and the environment may become non-negligible in a strong coupling regime. In such a regime, the equilibrium properties of the system depend on the interaction energy and the system state is no longer of the Gibbs form. Regarding such interactions, using  tools from the quantum estimation theory, we derive the thermodynamic uncertainty relation between intensive and extensive variables valid at all coupling regimes through the generalized Gibbs ensemble (GGE). Where we demonstrate the lower bound on the uncertainty of intensive variables increases in presence of quantum fluctuations. Also, we calculate the general uncertainty relations for several ensembles to corroborate the literature results, thus showing the versatility of our method.}

\vspace{1cm}
\end{abstract}

\vspace{1cm}

\newpage
 \renewcommand{\thefootnote}{*}
\section{Introduction}
{Traditional thermodynamics has always been concerned with describing macroscopic systems, while advances in the quantum regime suggest that microscopic systems may display unusual thermal behavior \cite{Talkner}. Quantum thermodynamics is an emerging research field based on the statistical mechanics approach to quantum systems. In statistical mechanics, statistical ensembles are used to analyze the thermodynamic systems at equilibrium. An ensemble describes a system that typically exchanges extensive quantities with its environment. Extensive quantities are variables that scale proportionally system size such as total volume or mass, in contrast to intensive variables that do not depend on system size such as temperature, or pressure.
Statistical ensembles are sometimes identified with the fixed natural variables from each statistically conjugated pair $(\beta, E)$, $(\mu, N)$, $(P, V)$ and so on \cite{Landau}. Where  $\beta=\frac{1}{k T}$ is the inverse temperature with the Boltzmann constant $k$ and $E$ is the energy, $P$ is the pressure and $V$ is the volume, $\mu$ is the chemical potential and $N$ is the number of particles. For example, in the canonical ensemble the system exchange only energy with the environment, and $\beta$ is an intensive parameter of the ensemble which is called $N V T$. The grand canonical ensemble exchange also particles, with the other intensive parameter $\mu$, and is called $\mu V T$. Moreover, the transition between the statistical ensembles is possible through Legendre transforms \cite{attard2002thermodynamics,tuckerman2010statistical,zia2009making,marzolino2020mu}.} 

{A crucial assumption in macroscopic thermodynamics is the weak coupling approximation, which states the equilibrium configuration of the system is not influenced by interactions with its environment. The equilibrium behavior of a macroscopic system describe by thermodynamic potentials such as internal energy $U$ as a function of the thermodynamic variables \cite{callen1985thermodynamics,Amaku}. Here, the term macroscopic refers to systems with a set of variables to guarantee the equilibrium properties in the thermodynamic limit. However, it is always possible to deviate from standard thermodynamics whenever we are dealing with the thermodynamics of open quantum systems, i.e., quantum systems that are allowed to interact with the environment. A recent research study has shown that in a strong coupling regime, the local equilibrium state of the system will not be of Gibbs form since the coupling strength between the system and environment influences the equilibrium properties of the system \cite{PhysRevE.86.061132}. Therefore, the thermodynamic potentials are substituted by the modified operators \cite{Philbin:1,Jarzynski:1,Strasberg:1}. In such a scenario, new energy fluctuations occur due to non-negligible coupling with the environment. In this case, the fluctuation of internal energy $U$ is determined via the fluctuation of a modified internal energy operator $E^{\ast}_{S}$ \cite{seifert2016first,miller2017entropy}. Regarding such unavoidable interactions, it is possible to derive an uncertainty relation between the internal energy and temperature of the system for arbitrary linear coupling scales by using quantum estimation theory that is valid for any coupling regime, where quantum fluctuations caused by coherence between the system energy states lead to higher fluctuations in the underlying temperature \cite{miller2018energy}. Moreover, it is not a secret that the study of statistical fluctuations caused by strong coupling effects can be very efficient for parameter estimation of nanoscale systems. Motivated by these considerations, we develop a general thermodynamic uncertainty relation via the quantum information measures for any pair of intensive and extensive thermodynamic variables for an arbitrary interaction strength in different ensembles. In fact, we find a generic expression of the uncertainty relations between Lagrange multipliers (intensive parameters) and their corresponding modified thermodynamic potentials, where quantum fluctuations increase the lower bound on the uncertainty of Lagrange multipliers. Our results are valid for all coupling regimes and are compatible with the particular case of the energy-temperature uncertainty in \cite{miller2018energy}.}

The paper is organized as follows. In Sec.~\ref{sec2}, we start with a generic approach towards the generalized Gibbs ensemble and the equilibrium entropy to pave the way where the skew and Fisher information are regarded as analytical tools. This approach fundamentally provides enough apparatus to access a generic extension to the strong coupling regime in other ensembles. Additionally, we display a brief overview of the toolkit of metrology quantities we used for our calculation and to derive our main results. The main results on the generic expression of uncertainty relations in the strong coupling regime are presented in Sec.~\ref{sec3}. We summarize our conclusions in Sec.~\ref{sec4}.

\section{Generic Approach}\label{sec2}
 To provide a general framework for uncertainty relation between each modified thermodynamic potential and its associated intensive parameter, it is essential to describe some important concepts.

At the start, a generalized Gibbs ensemble (GGE) is given by \cite{vidmar2016generalized, foini2017measuring, fukai2020noncommutative}
\begin{equation}
	\rho = \frac{e^{-\sum_i\lambda_i A_i}}{Z(\{\lambda_i\})}~,
\end{equation}
where $A_{i}$ and $\lambda_{i}$ are thermodynamic potentials and their corresponding Lagrange coefficients, respectively, and $Z(\{\lambda_i\})=\mathrm{tr}\left(e^{-\sum_i \lambda_iA_i}\right)$ is the partition function and, $[A_i,A_j]=0$ for all $i, j$. Hence, the expectation value of thermodynamic potential $A_{i}$ concerning the density matrix $\rho$ is $\langle A_i \rangle =-\frac{1}{Z}\frac{\partial Z}{\partial \lambda_i}$. Moreover, the statistical fluctuations for the constants of motion $A_i$ can be found via the partition function $Z$ as follows (see Appendix \ref{appA})
\begin{align}
	\frac{\partial^2 \ln Z}{\partial \lambda_i \, \partial \lambda_j} =\langle A_i A_j\rangle - \langle A_i\rangle\langle A_j\rangle~.
\end{align}
In this case, the natural logarithm of $Z$ is the Massieu function for the system \cite{swendsen2015continuity,swendsen2017resolving,swendsen2017thermodynamics}.

Also, it is straightforward to obtain the entropy of a system in thermodynamic equilibrium using the partition function in the following form
\begin{align}
	\frac{S}{k}=\ln Z - \sum_i\lambda_i\frac{\partial }{\partial \lambda_i}\ln Z~.
\end{align}
In addition, one can redefine $S$ from $\ln Z$ through a Legendre's transformation \cite{attard2002thermodynamics,tuckerman2010statistical,zia2009making,marzolino2020mu}. By considering infinitesimal changes $d\langle A_i \rangle$ and $d\lambda_i$ of $\langle A_i \rangle$ and $\lambda_i$ in equilibrium, respectively, we can obtain
\begin{align}
	d(\ln Z)&= - \sum_i \langle A_i \rangle d\lambda_i~, \nonumber\\
	dS &= k \sum_i\lambda_id\langle A_i \rangle ~.
\end{align}
It can be seen that  $d(\ln Z)$ and $dS$ capture the natural variables for $S$ and $\ln Z$, respectively. While $S$ is a function of $\langle A_i \rangle$ and $\ln Z$ is also beneficial as a function of $\lambda_i$, meaning that  (see Appendix \ref{appB} for more details)
\begin{align}
	\lambda_i &= \frac{1}{k}\frac{\partial S}{\partial \langle A_i\rangle}~, \nonumber\\
	\ln Z &= \frac{S}{k} - \frac{1}{k}\sum_i\langle A_i \rangle \frac{\partial  S}{\partial\langle A_i \rangle } ~.
\end{align}
In probability and statistics theory, the variance of an operator is defined as
\begin{align}
	\mathrm{Var}(\rho,\mathcal{O})=\mathrm{tr}(\rho\,\mathcal{O}^2)-[\mathrm{tr}(\rho\,\mathcal{O})]^2~,
\end{align}
for each state $\rho$ and observable $\mathcal{O}$. Two important features of this quantity are positivity and concavity \cite{zhang2021tighter}. Moreover, in the context of quantum measurements, Wigner and Yanase introduced the information measure so-called the Wigner-Yanase skew information (WYSI), which is determined by $Q_{1/2}({\rho}, \mathcal{O})=-\frac{1}{2}{\mathrm{tr}([\sqrt{\rho}, \mathcal{O}]^{2})}$ \cite{zhang2021tighter,takagi2019skew}. It is proposed to quantify the information due to non-commutativity between the state $\rho$ and the observable $\mathcal{O}$ \cite{pires2020relating}. WYSI is also generalized by Dyson as 
$Q_\alpha({\rho},\mathcal{O})=-\frac{1}{2}tr([\mathcal{O} ,\rho^{\alpha}][\mathcal{O} ,\rho^{\alpha-1}])$ , which is called Wigner-Yanase-Dyson skew information (WYDSI) \cite{luo2003wigner}. Moreover, the complementary measure of classical uncertainty is given by $K_\alpha({\rho},\mathcal{O})=tr(\rho^{\alpha}\delta \mathcal{O}\rho^{\alpha-1}\delta \mathcal{O})$ with $\delta \mathcal{O}=\mathcal{O} - \langle \mathcal{O}\rangle$ and $\alpha\in(0 , 1)$ \cite{luo2005heisenberg}. With parameter $\alpha$, it is impossible to separate the quantum and classical contributions to variance in a unique way. However, this issue can be resolved by averaging over $\alpha$ for the entire interval, where the quantum and classical uncertainty contributions of quantum observable $\mathcal{O}$ in a mixed state $\rho$ are separated by the following expressions \cite{frerot2016quantum} 
\begin{align}
	Q(\rho , \mathcal{O})&=\int_{0}^{1} d\alpha~ Q_{\alpha}(\rho , \mathcal{O}),\\
	K(\rho , \mathcal{O})&=\int_{0}^{1} d\alpha~ K_{\alpha}(\rho , \mathcal{O})~.
\end{align}
The skew information has several interpretations, such as a version of the quantum Fisher information (QFI) \cite{luo2003wigner,luo2004wigner,luo2012quantifying}, quantum uncertainty of $\mathcal{O}$ on state $\rho$ \cite{luo2012quantifying,luo2005heisenberg,luo2007brukner}, coherence, and asymmetry of $\rho$ concerning $\mathcal{O}$  \cite{girolami2013characterizing,sun2017quantum,luo2019quantifying,marvian2016quantum,luo2018coherence}. All of these imply that skew information is a significant concept. However, in the case of mixed state $\rho$, it has been reported that $\mbox{Var}(\rho,\mathcal{O})\ge{Q(\rho,\mathcal{O})}$ \cite{luo2003wigner,luo2006quantum}. Furthermore, some substantial features of skew information include:

\begin{itemize}
	\item $Q(\rho,\mathcal{O})$ is non-negative and equal to the variance of observable $\mathcal{O}$ for a pure state.
	
	\item It remains constant for isolated systems.
	
	\item $Q(\rho,\mathcal{O})$ is a convex function because the mixing process of states is entirely classical and cannot increase quantum uncertainty.
	
	\item It is also additive, i.e., the information of a system composed of two independent parts is equal to the sum of the information of each part separately.
\end{itemize}
In the following, we consider
\begin{equation}\label{S}
	\rho = \frac{e^{-\sum_i \lambda_i A_i}}{Z(\{\lambda_i\})}=e^G 
\end{equation}
where $G=-\sum_i \lambda_iA_i -\ln Z$ and $[A_i,\rho]=0$. Hence $K_\alpha(\rho,G)=\mathrm{Var}[\rho,G]$ and also $Q_\alpha(\rho,G)=0$. According to \cite{luo2005heisenberg}, when $A_i$ and $\rho$ commute, quantum uncertainty based on the skew information should vanish.

In the context of quantum-enhanced parameter estimation, a suitable measure to determine the ultimate limit of estimation precision is the Cram\'er-Rao (CR) bound \cite{fanchini2017lectures,cavina2018bridging,liu2019quantum}
\begin{equation}
	\Delta{\boldsymbol{\theta}}\geq{\frac{1}{\sqrt{n\mathcal{F}(\theta)}}}~,
\end{equation}
with $\mathcal{F}(\theta)$ being the QFI \cite{giovannetti2011advances} and $\theta$ standing for the unknown parameter encoded in the initial state. The authors in \cite{miller2018energy} prove that for a quantum exponential state $\rho_{\theta}=e^{- \mathcal{O}_{\theta}}/Z_{\theta}$, where $\mathcal{O}_{\theta}$ is a Hermitian operator and $Z_{\theta}=tr(e^{- \mathcal{O}_{\theta}})$, the QFI $\mathcal F(\theta)$ concerning the parameter $\theta$ is bounded as follows (see theorem $1$ in \cite{miller2018energy})
\begin{align}\label{F}
	\mathcal F(\theta)\leq K(\rho_{\theta},B_{\theta}),
\end{align}
with $B_{\theta}=\partial_{\theta}\mathcal{O}_{\theta}$ being a Hermitian observable. Equation \eqref{F} displays that the achievable precision in estimating parameter $\theta$ is constrained by the strict fluctuations in $B_{\theta}$.

As before, by considering $\rho = e^G$ with $G= -\sum_i\lambda_iA_i - \ln Z$ such that $G\,|e_j\rangle = g_j\,|e_j\rangle$, we can write the QFI in the following form \cite{modi2011quantum,hayashi2016quantum,fiderer2019maximal}
\begin{equation}
	\mathcal F(\lambda_p)=2\sum_{n,m}\frac{|\langle e_n|\partial_{\lambda_p}\rho|e_m\rangle|^2}{p_n+p_m}~,
\end{equation}
for the desired Lagrange coefficient $\lambda_{p}$. Here, $A_n\,|e_m\rangle = a_{nm}\,|e_m\rangle$, and $\rho=\sum_n p_n |e_n\rangle\langle e_n|$ is the spectral decomposition of $\rho$.
Therefore, by rewriting the GGE state in Eq.~\eqref{S} as
\begin{align}
	\rho =\sum_ne^{G}|e_n\rangle\langle e_n|=\sum_ne^{g_n}|e_n\rangle\langle e_n|~,
\end{align}
and substituting $g_n=\ln(p_n)$, it turns out that (see Appendix \ref{appC})
\begin{equation}\label{14}
	\mathcal F(\lambda_p) = \sum_n p_n (a_{pn}-\langle A_p\rangle )^2~.
\end{equation}
At this point, by using the fact that $K_\alpha(\rho,G)= \mathrm{Var}(\rho,G)=\sum_n\langle e_n| e^G (\delta G)^2 |e_n\rangle$, we obtain (see Appendix \ref{appC})
\begin{align}\label{15}
	\mathrm{Var}(\rho,G)=\sum_{n,i}p_n \lambda_i^2(a_{ni}-\langle A_i \rangle )^2~.
\end{align}
As a result, because the WYDSI is zero, a fair comparison occurs when the sum of the QFI's for each parameter $\lambda_i$ coincides with  $\mathrm{Var}[\rho,G]$, which is consistent with the results of \cite{miller2018energy}. 

Nevertheless, when $A_{i}$ is a function of $\lambda_i$, condition $[A_{i},\rho]=0$ is no longer satisfied and subsequently $Q_{\alpha}(\rho,G)\neq 0$. On the one hand, the term $Q_{\alpha}(\rho,G)$ stems from the dependence between operator $A_{i}$ and Lagrange coefficient $\lambda_i$. On the other hand, it cannot be zero where there is a strong interaction strength between the system and the environment. Therefore, the crucial point distinguishing a strong coupling regime from a weak one is the dependence between $A_{i}$ and $\lambda_i$. In other words, when the interaction energy between the system and its environment is not weak enough, the contributions of $A_{i}$ and $\lambda_i$ to the reduced density matrix of the system after partial trace over the environment's degrees of freedom will not be independent.

\section{Main Results}\label{sec3}
\subsection{Strong Coupling Approach}
From now on, we provide a generic expression of the uncertainty relations between Lagrange coefficients and their corresponding thermodynamic potentials in the strong coupling regime. To this end, we assume the quantum system $S$ interacts strongly with the environment $E$ by the Hamiltonian
\begin{equation}
	H_{SE}: = H_S + H_E + H_{\text{int}}~,
\end{equation}
where $H_S$ and $H_E$ are the Hamiltonians of the system and the environment, respectively, and $H_{\text{int}}$ is an interaction Hamiltonian that captures the arbitrary interaction strength between $S$ and $E$. Let us consider the state of the composite system as the generalized Gibbs form
\begin{equation}
	\rho_{SE} = \frac{e^{-\sum_{k} \lambda_k I_k}}{Z_{SE}}~,
\end{equation}
where $I_{k}$'s are thermodynamic potentials of the generalized Gibbs state of the total system ($S+E$), and $[I_k,I_{k^{\prime}}]=0$. 

In a particular case where a quantum system $S$ strongly interacts with a
canonical reservoir $E$, the state of the composite system at \textcolor{red}{inverse} temperature $\beta$ has a Gibbs form $\rho_{SE}(\beta)=e^{-\beta H_{SE}}/Z_{SE}$, where the partition function of the system and the environment is given by $Z_{SE}=tr(e^{-\beta H_{SE}})$. The reduced state of the system $S$ is characterized by $\rho_{S}(\beta)=tr_{E}(\rho_{SE}(\beta))$ and is generally not thermal concerning $H_{S}$ due to the non-negligible interaction term. Thus, the partition function that is denoted by $H_S$ would not be adequate to calculate the internal energy of the system \cite{seifert2016first}. However, one can easily overcome this issue by defining an effective Gibbs state $\rho_{S}(\beta)=e^{-\beta H^{\ast}_{S}(\beta)}/Z^{\ast}_{S}$, where $H^{\ast}_{S}(\beta)=\frac{1}{\beta}\ln(tr_{E}(e^{-\beta H_{SE}})/\mathrm{tr}_{E}(e^{-\beta H_{E}}))$ is the Hamiltonian of mean force (HMF), and $Z^{\ast}_{S}=\mathrm{tr}_{S}(e^{-\beta H^{\ast}_{S}(\beta)})$ is the effective partition function \cite{jarzynski2004nonequilibrium,hilt2011hamiltonian,strasberg2017stochastic}. 

In this sense and according to the results of the previous section, one can conclude that in the strong coupling regime where the interaction energy between the system and the environment is not negligible, the reduced density matrix $\rho_{S}$ is no longer of the Gibbs form. It can generically be written as 
\begin{equation}\label{rho}
	\rho_S=\frac{e^{-\sum_i\lambda_iA^*_i}}{Z_S^*}~,
\end{equation}
where the effective thermodynamic potentials $A^{\ast}_i$'s of the system typically are complicated operator-valued functions of the system's parameters and other environmental parameters including the system-environment coupling strength, the environment temperature and etc. In a weak coupling regime, the contribution of the environment and interaction is negligible, and the effective thermodynamic potentials are reduced to those of the bare system.

Here, the reduced density state $\rho_{S}$ in Eq.~\eqref{rho} is known as \textit{the generalized effective Gibbs state} of the system. $Z^*_S$ can be defined via the normalization condition $Z_S^*=\frac{Z_{SE}}{Z_E}$ and $A_i^*$ is obtained through
\begin{align}
		\sum_i\lambda_i A_i^* = -\ln \left(Z_S^*\mathrm{tr}_E(\rho_{SE})\right)
		=-\ln\left\{\frac{\mathrm{tr}_E(e^{-\sum_k\lambda_kI_k})}{Z_E}\right\}~,
\end{align}
where the partition function of the environment is given by $Z_E=\mathrm{tr}\textcolor{red}{(}e^{-\sum_j\lambda_j R_j}\textcolor{red}{)}$ with $[R_j,H_E] =0$; one can also define $Z_S^*=\mathrm{tr}\textcolor{red}{(}e^{-\sum_i\lambda_iA_i^*}\textcolor{red}{)}$. In what follows, we present a generic thermodynamic potential for this situation. Due to this fact that $A^*_i$ depends on $\lambda_i$, one can write \cite{seifert2016first} 
\begin{equation}
	\biggl\langle \frac{\partial}{\partial{\lambda_p}}\sum_i\lambda_i{A^*_i}\biggr\rangle =\langle {E^*_S}\rangle = \Phi_s~.
\end{equation}
The left-hand side of the above equation can be calculated in terms of the partition function, so we have
\begin{align}
	\biggl\langle \frac{\partial}{\partial{\lambda_p}}\sum_i\lambda_i{A^*_i}\biggr\rangle =- \frac{\partial}{\partial \lambda_p}(\ln Z_S^*)~.
\end{align}
To derive a universal uncertainty relation for two thermodynamically conjugated quantities, we assume that ${\rho_S}=e^{G_S^*}$, in which $G_S^*= -\sum_i\lambda_iA^*_i - \ln Z_S^*$. According to the previous methods, the QFI would be given by (see Appendix \ref{appD})
\begin{equation}\label{21}
	\mathcal F(\lambda_p)=\mathrm{Var}(\rho_S,E^*_S)+\Xi~,
\end{equation}
where
\begin{equation}
	\Xi=\sum_{n\neq m}\left[\frac{2(p_m-p_n)^2}{(p_n+p_m)(\ln(p_m/p_n))^2}-p_n\right]|\langle e_n| E^*_S |e_m\rangle|^2~.
\end{equation}
At this point, one can obtain (see Appendix \ref{appE})
\begin{equation}\label{23}
	Q(\rho_S,E^*_S)=\sum_{n\neq m}\left(p_n-\frac{p_m-p_n}{\ln(p_m/p_n)}\right)|\langle e_n|E^*_S|e_m\rangle|^2~.
\end{equation}
By adding the terms related to the skew information \eqref{23} to the Fisher information \eqref{21} and taking into account the considerations presented in Appendix \ref{appF} (please follow the derivation process from Eq.~\eqref{72} to Eq.~\eqref{F6}), we can write an inequality as the following form
\begin{equation}\label{24}
	\mathcal F(\lambda_p) \leq \mathrm{Var}(\rho_S,E^*_S)-Q(\rho_S,E^*_S)~.
\end{equation}
Here, we use the Cramer-Rao bound in a single-shot scenario ($n=1$), which is given by 
\begin{equation}
	\Delta \theta \geq \frac{1}{\sqrt{\mathcal F(\theta)}}~.
\end{equation}
Thus, for $\theta=\lambda_p$, we find that
\begin{equation}\label{p}
	\Delta {\lambda_p} \geq \frac{1}{\sqrt{\mathrm{Var}(\rho_S,E^*_S)-Q(\rho_S,E^*_S)}}~,
\end{equation}
and one can prove 
\begin{align}
	\mathrm{Var}(\rho_S ,E^*_S)-Q(\rho_S,E^*_S)&= K(\rho_S,E^*_S)~.
\end{align}
Furthermore, we can express the uncertainty using the operators $A^{*}_i$'s explicitly by
\begin{equation}
	{E^{*p}_{S}}=\sum_i\frac{\partial(\lambda_iA^*_i)}{\partial\lambda_p}~.
\end{equation}
The Lagrange coefficients are usually chosen to correspond to thermodynamic observables, and more precisely to entropic-intensive parameters. This simplifies the translation between the distributions and thermodynamic relations. Moreover, we assume that each conserved quantity is characterized by a unique Lagrange coefficient \cite{WACHSMUTH201378}. By considering $\lambda_i$ and $A^*_p$ dependence on $\lambda_p$, we have
\begin{align}
	E^{*p}_{S} = \sum_{i\neq p}(\dot{\lambda}_iA_i^*+\lambda_i\dot{\lambda}_i\frac{\partial A_i^*}{\partial \lambda_i})+\partial_{\lambda_p}[\lambda_p A_p^*]~,
\end{align}
where $\dot{\lambda}_i=\frac{\partial \lambda_i}{\partial \lambda_p}$. Given that $[A^*_p,\rho_S]=0~(p\neq i)$, we derive a general expression for classical uncertainty as (see Appendix \ref{appF})
\begin{align}\label{29}
	K(\rho_S, E^{*p}_{S})&=K(\rho_S,\partial_{\lambda_p}[\lambda_p A^*_p])+2\sum_{i\neq p} \dot{\lambda}_i~ \mathrm{Cov}~(\partial_{\lambda_p}[\lambda_p A^*_p],~A^*_i+\lambda_i \frac{\partial A^*_i}{\partial\lambda_i})\nonumber\\&+\sum_{ij\neq p}\dot{\lambda}_i\dot{\lambda}_j~ \mathrm{Cov}~(A^*_i+\lambda_i \frac{\partial A^*_i}{\partial\lambda_i},~A^*_j+\lambda_j\frac{\partial A^*_j}{\partial\lambda_j})~.
\end{align}
\textcolor{red}{where $\mathrm{Cov}(X,Y)=\langle X Y\rangle-\langle X\rangle \langle Y\rangle$.} If $A_i^*$ and $A_j^*$ commute for $i,j \neq p$, the last covariance term between them vanishes. Also, it can be seen in general that the Lagrange coefficient changes the sign with the derivative. The minus sign can appear for some specific examples in the following examples.

In the next step, we investigate the general uncertainty relation in different statistical ensembles and acquire the uncertainty between the characteristic parameters of each one. According to statistical mechanics, each ensemble is described by a physical system that exchanges special physical quantities in interaction with its environment. Furthermore, statistical ensembles are labeled by variables that remain constant from each pair of statistical conjugates, such as $(\beta ,E), (\mu, N)$, and $(P, V)$, etc.
\subsubsection{Canonical Ensemble}
In the canonical ensemble (CE), the system exchanges only energy with the environment.  Thermodynamic control variables characterize the CE: constant temperature $T$, constant volume $V$, and fixed particle number $N$, or ($TVN$). The system state is defined in the following form 
\begin{equation}\label{o}
	\rho_S = e^{-\sum_i\lambda_i{A_i^{*}}-\ln Z_S^*}= e^{-\beta H^*_S-\ln Z_S^*}~.
\end{equation}
Since the sum is reduced to a single element in Eq.\,\eqref{o}, one can take $\lambda_p = \beta$, so $E^*_S = \partial_{\beta} (\beta H^*_S)$ as \cite{miller2018energy} and the thermodynamic potential is given by ${U}_S=\langle {E^*_S} \rangle$. We recover the result in \cite{miller2018energy}, which reads as follows
\begin{equation}\label{30}
	\Delta \beta \geq \frac{1}{\sqrt{\Delta U_S^{2} - Q(\rho_S ,E^*_S)}} \geq \frac{1}{\Delta U_S}~.
\end{equation}
According to the uncertainty relation \eqref{30}, quantum fluctuations arising from coherence between energy states will increase the temperature fluctuation at a given spread in energy \cite{miller2018energy}. In this case, the additional temperature fluctuations are measured by the averaged WYDSI.

\subsubsection{Grand Canonical Ensemble}

In the grand canonical ensemble (GCE), the system is permitted to exchange particles and energy with the environment. Thermodynamic control variables characterize the GCE: chemical potential $\mu$, constant volume $V$, constant temperature $T$, or ($\mu VT$). In the present case, we have
\begin{equation}
	\rho_S = e^{-\sum_i\lambda_i{A_i^*}-\ln Z_S^*}= e^{-\beta(H_S^*-\mu N^{*}_S)-\ln Z_S^*}~,
\end{equation}
which in turn gives $\sum_i \lambda_i A_i^* = \beta(H_S^* - \mu N^{*}_S)$, so the intensive quantities are defined as 
\begin{equation}
	\lambda_1 = \beta \qquad \lambda_2 = -\beta\mu~.
\end{equation}
Now, it is feasible to calculate the uncertainty relation for the temperature and internal energy, where we consider $\lambda_1 = \beta$ and $\lambda_2 = -\mu\lambda_1$. One can write
\begin{align}
	{E^{*(p=1)}_S} &= \frac{\partial}{\partial\lambda_1}\left(\sum_i \lambda_i A^*_i\right)=\left(\dot{\lambda_2}N_S^{*} + \lambda_2\dot{\lambda_2}\frac{\partial N_S^*}{\partial\lambda_2}\right)+\partial_{\lambda_1}[\lambda_1 H_S^*]\nonumber\\&=\mathcal{A^{*}}-\mu\mathcal{B^{*}}~,
\end{align}
where $\dot{\lambda_2}=\frac{\partial\lambda_2}{\partial\lambda_1}$, ~$\mathcal{A^{*}}=\partial_{\lambda_1}[\lambda_1 H^*_S]$, and $\mathcal{B^{*}}=N_S^{*}+\lambda_2\frac{\partial N_S^{*}}{\partial\lambda_2}$. Thus
\begin{align}
	\langle {E^{*(p=1)}_S} \rangle = \mathrm{tr}[\rho_{S}(\mathcal{A^{*}}-\mu\mathcal{B^{*}})]
	= \langle \mathcal{A^{*}}\rangle - \mu \langle\mathcal{B^{*}}\rangle~.
\end{align}
According to the above-established definition
\begin{align}\label{k}
	K(\rho_S,\mathcal{A^{*}}-\mu\mathcal{B^{*}})=K(\rho_S , \mathcal{A^{*}})+\mu^2\mathrm{Var}(\rho_S,\mathcal{B^{*}})
	-2\mu\mathrm{Cov}(\mathcal{A^{*}},\mathcal{B^{*}})~.
\end{align}
The upper bound obtained on the QFI is as follows
\begin{align}
	\mathcal{F}(\lambda_1)\leq K(\rho_S,\mathcal{A^{*}})+\mu^2\mathrm{Var}(\rho_S,\mathcal{B^{*}})-2\mu\mathrm{Cov}(\mathcal{A^{*}},\mathcal{B^{*}})~.
\end{align}
Regarding Eq.~\eqref{24}, we have
\begin{align}
	\Delta\lambda_1\geq\frac{1}{\sqrt{K(\rho_S,\mathcal{A^{*}})+\mu^2\mathrm{Var}(\rho_S,\mathcal{B^{*}})-2\mu\mathrm{Cov}(\mathcal{A^{*}},\mathcal{B^{*}})}}~.
\end{align}
Moreover, by definition $\Delta\Phi_{S}=\sqrt{\mathrm{Var}(\rho_{S},~\mathcal{A^{*}}-\mu\mathcal{B^{*}})}$,

\begin{align}
	\mathrm{Var}(aX\pm bY)=a^{2}\mathrm{Var}(X)+b^{2}\mathrm{Var}(Y)\pm 2ab~\mathrm{Cov}(X,Y)~,
\end{align}
and since the skew information is positive and also the covariance elements are positive semi-definite, one can write
\begin{align}
	\Delta\lambda_1\geq\frac{1}{\sqrt{\Delta\Phi_{S}^{2}- Q(\rho_S,\mathcal{A^{*}})}}\geq\frac{1}{\Delta\Phi_{S}}~.
\end{align}
To find the uncertainty relation between $\lambda_2=-\beta\mu$ and $N_S^{*}$, one can redefine $\lambda_1=-\lambda_2/\mu$, and then write
\begin{align}
	{E^{*(p=2)}_S} &= \frac{\partial}{\partial\lambda_2}\left(\sum_i \lambda_i A^*_i\right)=\left(\dot{\lambda_1}H_S^{*} + \lambda_1\dot{\lambda_1}\frac{\partial H_S^*}{\partial\lambda_1}\right)+\partial_{\lambda_2}[\lambda_2 N_S^*]\nonumber\\&=\mathcal{B^{*}}-\frac{1}{\mu}\mathcal{A^{*}}~,
\end{align}
where $\dot{\lambda_1}=\frac{\partial\lambda_1}{\partial\lambda_2}$, hence 
\begin{align}
	\langle {E^{*(p=2)}_S} \rangle=\mathrm{tr}[\rho_{S}(\mathcal{B^{*}}&-\frac{1}{\mu}\mathcal{A^{*}})]
	= \langle\mathcal{B^{*}}\rangle -\frac{1}{\mu}\langle\mathcal{A^{*}}\rangle~,\\
	K(\rho_S,\mathcal{B^{*}}-\frac{1}{\mu}\mathcal{A^{*}}))&=K(\rho_S , \mathcal{B^{*}})+\mu^{-2}~\mathrm{Var}(\rho_S,\mathcal{A^{*}})\nonumber\\
	&-2\mu^{-1}\mathrm{Cov}(\mathcal{B^{*}},\mathcal{A^{*}})~,
\end{align}
and the upper bound obtained on the QFI is
\begin{align}
	\mathcal F(\lambda_2)\leq K(\rho_S,\mathcal{B^{*}})+\mu^{-2}\mathrm{Var}(\rho_S,\mathcal{A^{*}})-2\mu^{-1}\mathrm{Cov}(\mathcal{B^{*}},\mathcal{A^{*}})~.
\end{align}
Subsequently, the uncertainty relation for the intensive parameter $\lambda_2$ turns into
\begin{align}
	\Delta\lambda_2\geq  \frac{1}{\sqrt{K(\rho_S,\mathcal{B^{*}})+\mu^{-2}\mathrm{Var}(\rho_S,\mathcal{A^{*}})-2\mu^{-1}\mathrm{Cov}(\mathcal{B^{*}},\mathcal{A^{*}})}}~.
\end{align}
Given that $\Delta \Phi_S=\sqrt{\mathrm{Var}(\rho_S,\mathcal{B^{*}}-\mathcal{A^{*}}/\mu)}$, and with a similar argument to the previous case, we have
\begin{align}\label{42}
	&\Delta\lambda_2\geq \frac{1}{\sqrt{\Delta \Phi_S^2-Q(\rho_S,\mathcal{B^{*}})}}\geq \frac{1}{\Delta\Phi_S}~.
\end{align}

\subsubsection{Other Ensembles}
Attractively, our approach applies to cases with more than two Lagrange coefficients in the strong coupling regime. The following density operators include the systems of interest: \cite{callen1985thermodynamics}
\begin{itemize}
	\item The ensemble of mental copies of the system that moves freely in space and is given by
	\begin{align}
		\rho = \frac{1}{Z}e^{-\beta H-\lambda_xP_x-\lambda_yP_y-\lambda_zP_z}~.
	\end{align}
	\item Magnetic substances with the Hamiltonian $H_0$ when there is no field. We try to study their properties as a function of the total magnetic moment in the following form 
	\begin{align}
		\rho = \frac{1}{Z}e^{-\beta H_0 +\beta \vec{B} \cdot \vec{M}}~.
	\end{align}
	The data are $\langle H_0\rangle$, $\langle M_{i} \rangle$ with $i=x,y,z$; and the Lagrange coefficient $-\beta B_{i}$ can be identified apart from the factor $-\beta$ as an external magnetic induction necessary to produce the moment $\langle M_{i} \rangle$.
	\item With the total angular momentum $\vec{J}$ and the Hamiltonian $H_0$ in which the system does not rotate, we have \cite{Halpern:1,Halpern:2}
	\begin{align}
		\rho =\frac{1}{Z}e^{-\beta H_0 + \beta \vec{\omega} \cdot \vec{J}}~,
	\end{align}
	where $\vec{\omega}$ is an angular velocity.
\end{itemize}
The above cases need four Lagrange coefficients and must be given by four different uncertainty relations. We may derive them more generically by density operator
\begin{align}
	\rho_S = \frac{1}{Z_S^*}e^{-(\beta H^*_S -\beta\sum_{i=1}^{3}\mu_iI_i)}~.
\end{align}
In this case, each $\mu_i$ is an intensive property with the associated operator $I_i$ of the system, and $H_S^*$ is the HMF. Therefore, they commute with the density matrix. In our approach, the Lagrange coefficients are given by
\begin{align}
	\lambda_0 = \beta, \quad \lambda_1 = -\beta\mu_1, \quad \lambda_2 = -\beta\mu_2, \quad \lambda_3 = -\beta\mu_3 ~.
\end{align}
As an example, to calculate the uncertainty relation between $H_S^{*}$ and its corresponding intensive parameter $\lambda_0 =\beta$, we can rewrite the other coefficients as $\lambda_1 = -\mu_1\lambda_0,~\lambda_2 = -\mu_2\lambda_0$, and $\lambda_3 = -\mu_3\lambda_0$. So, by formulating
\begin{align}
	E^{*(p=0)}_S &=\sum_{i=1}^{3}(\dot{\lambda_i}I_i^{*}+\lambda_i\dot{\lambda_i}\frac{\partial I_i^{*}}{\partial \lambda_i})+\partial_{\lambda_0}[\lambda_0 H_S^{*}]\nonumber\\&=\mathcal{D^{*}}-\sum_{i=1}^{3}\mu_{i}\mathcal{I}^{*}_{i}~,
\end{align}
where $\mathcal{D^{*}}=\partial_{\lambda_0}[\lambda_0 H_S^{*}]$, $\dot{\lambda_i}=\frac{\partial\lambda_i}{\partial\lambda_0}$ and $\mathcal{I}^{*}_{i}=I_i^{*}+\lambda_i\frac{\partial I_i^{*}}{\partial\lambda_i}$, we have
\begin{align}\label{53}
	K(\rho_S,{E^{*(p=0)}_S})=K(\rho_S,\mathcal{D}^*)-2\sum_{i=1}^{3}\mu_i\mathrm{Cov}(\mathcal{D}^*,\mathcal{I}^{*}_{i})+\mathrm{Var}(\rho_{S},\sum_{i=1}^{3}\mu_{i}\mathcal{I}^{*}_{i})~.
\end{align}

To derive the last term on the right side of Eq.~\eqref{53}, we use the fact that $\mathrm{Var}(\sum_{i}a_{i}X_{i})=\sum_{i,j}a_{i}a_{j}\mathrm{Cov}(X_{i},X_{j})$. At this point, the uncertainty relation is obtained as
\begin{align}
	\Delta\lambda_0\geq \frac{1}{\sqrt{K(\rho_S,\mathcal{D}^*)-2\sum_{i=1}^{3}\mu_i\mathrm{Cov}(\mathcal{D}^*,\mathcal{I}^{*}_{i})+\mathrm{Var}(\rho_{S},\sum_{i=1}^{3}\mu_{i}\mathcal{I}^{*}_{i})}}~.
\end{align}
By definition $\Delta \Phi_S = \sqrt{\mathrm{Var}~(\rho_S,\mathcal{D}^* -\sum_{i=1}^{3}\mu_i \mathcal{I}^{*}_i)}$ and taking an approach similar to that of the previous section, we arrive at
\begin{align}\label{54}
	\Delta\lambda_0\geq \sqrt{\Delta \Phi_S^{2} -Q(\rho_S,\mathcal{D}^*)}\geq \frac{1}{\Delta \Phi_S}~.
\end{align}
As a main result, we can deduce from Eqs.~ \eqref{30}, \eqref{42}, and \eqref{54} that the inclusion of quantum effects yields a tighter lower bound. In other words, quantum fluctuations increase the lower bound on the uncertainty of the environment's intensive parameters. Note that the quantum uncertainty contribution vanishes for classical systems in the equations. Indeed, the present approach via the quantum information measures such as the QFI and the average WYDSI makes a general methodology to investigate the quantum fluctuation effects compared with the classical ones in quantum thermodynamics.
In addition, our results indicate that in a strong coupling regime, the uncertainty bounds are affected by the correlations between the effective thermodynamic potentials of the system and the intensive parameters of the environment. Therefore, it is necessary to determine the effective potentials for any thermodynamic system before implementing these bounds. In this sense, our framework compares standard thermodynamics with sufficiently weak interactions and cases where the correlations play a significant role. So, a detailed analysis of our approach in different thermodynamic systems can lead to novel and exciting results in the future.

\section{Conclusion}\label{sec4}
The use of thermodynamics in quantum regimes has recently gained interest, where tiny microscopic systems are used in the laboratory for high-precision measurements. Quantum thermodynamics has attracted considerable attention because of its potential applications in numerous technological fields. However, as the size of the systems under study grows smaller, the fluctuations of thermodynamic variables are more important during the system measurement and manipulation. Moreover, in the thermodynamics realm of open quantum systems, unavoidable environmental interactions affect the uncertainty of thermodynamic variables. Therefore, it is crucial in quantum thermodynamics to calculate these fluctuations and determine the precision bounds on them. Here, we derived various forms of interesting general uncertainty relations, which are helpful for estimating thermodynamic variables. We provided a general framework for calculating the uncertainty relation for each pair of thermodynamic intensive and extensive variables. To this end, we regarded the transition approach between different ensembles at the thermodynamic limit via a Legendre transformation \cite{attard2002thermodynamics,tuckerman2010statistical,zia2009making,marzolino2020mu}. It should be mentioned that the average WYDSI evaluates the fluctuation of intensive variables. Finally, we found a generalized and valid uncertainty relation between intensive and extensive variables for all coupling regimes, using the QFI constraint for exponential states. Also, to illustrate the comprehensiveness and features captured by this set of uncertainties, we used several ensembles.
Our results indicate that uncertainty bounds for intensive quantities increase whenever quantum fluctuations occur in thermodynamic potentials. These additional fluctuations lead to a tighter lower bound and more accurate parameter estimation. In addition, it is fair to claim that the results presented in \cite{miller2018energy} can be considered as a specific case of the general framework presented here.


\section*{Data availability}
No datasets were generated or analyzed during the current study.


\section*{Competing interests}
The authors declare no competing interests.

\appendix


\section{Partition Function}\label{appA}
The generalized Gibbs ensemble (GGE) is given by \cite{vidmar2016generalized, foini2017measuring, fukai2020noncommutative}
\begin{equation}
	\rho = \frac{e^{-\sum_i\lambda_iA_i}}{Z(\{\lambda_i\})}~,
\end{equation}
where $[A_i,A_j]=0$ for all $i, j$, and the partition function is
\begin{equation}
	Z(\{\lambda_i\})=\mathrm{tr}\left(e^{-\sum_i \lambda_iA_i}\right)~,
\end{equation}
a situation in which the ensemble is in thermodynamic equilibrium. It is important to note that
\begin{align}
	\langle A_i \rangle &= \mathrm{tr}\left(\rho A_i\right)=\frac{1}{Z}\mathrm{tr}\left(e^{-\sum_i\lambda_iA_i}A_i\right)\nonumber\\&=-\frac{1}{Z}\frac{\partial }{\partial \lambda_i}\mathrm{tr}\left(e^{-\sum_i \lambda_i A_i}\right)~,
\end{align}
consequently
\begin{equation}\label{eq:average-A-by-part.function}
	\langle A_i \rangle =-\frac{\partial}{\partial \lambda_i}\ln Z(\{\lambda_i\})=-\frac{1}{Z}\frac{\partial Z}{\partial \lambda_i}~.
\end{equation}
Moreover, the statistical fluctuations for the constants of motion $A_i$ can be found via the partition function $Z$ as follows
\begin{align}\label{z}
	\frac{\partial^2 Z}{\partial \lambda_i \, \partial \lambda_j}=\mathrm{tr}\left(e^{-\sum_k\lambda_kA_k}A_i A_j\right)~.
\end{align}
However, the calculations become easier when $Z$ is replaced by $\ln Z$ in Eq.\, \eqref{z}, then we have
\begin{align}
	\frac{\partial^2 \ln Z}{\partial \lambda_i \, \partial \lambda_j} &=\frac{\partial }{\partial \lambda_i}\frac{1}{Z}\mathrm{tr}\left(e^{-\sum_k\lambda_kA_k}A_j\right)=\frac{1}{Z^2}\frac{\partial Z}{\partial \lambda_i}Z\langle A_j\rangle+\frac{1}{Z}\mathrm{tr}\left(e^{-\sum_k\lambda_kA_k}A_iA_j\right)\nonumber\\
	&=\langle A_iA_j\rangle - \langle A_i\rangle\langle A_j\rangle ~.
\end{align}
\section{Equilibrium Entropy}\label{appB}
It is straightforward to obtain the entropy of a system in thermodynamic equilibrium via the partition function in the following path
\begin{align}
	\frac{S}{k}=-\langle \ln \rho \rangle &= - \mathrm{tr}(\rho\ln\rho)\nonumber\\
	&=-\mathrm{tr}\left\{\frac{e^{-\sum_i\lambda_iA_i}}{Z}\ln\left(\frac{e^{-\sum_i\lambda_iA_i}}{Z}\right)\right\}\nonumber\\
	&=\ln Z \,\mathrm{tr}\left\{\frac{e^{-\sum_i\lambda_iA_i}}{Z}\right\}+\sum_i\lambda_i\,\mathrm{tr}\left\{\frac{e^{-\sum_i\lambda_iA_i}}{Z}A_i\right\}\nonumber\\
	&=\ln Z - \sum_i\lambda_i\frac{\partial }{\partial \lambda_i}\ln Z ~.
\end{align}
Moreover, one can redefine $S$ from $\ln Z$ by a Legendre transformation \cite{attard2002thermodynamics,tuckerman2010statistical,zia2009making,marzolino2020mu}. Let us consider the infinitesimal changes $d\langle A_i \rangle$ and $d\lambda_i$ of $\langle A_i \rangle$ and $\lambda_i$ in equilibrium, respectively. Accordingly,
\begin{equation}\label{eq:differential of lnZ}
	d(\ln Z)= \sum_i\frac{\partial \ln Z}{\partial \lambda_i}d\lambda_i = - \sum_i \langle A_i \rangle d\lambda_i ~,
\end{equation}
and
\begin{equation}
	\frac{dS}{k}=d(\ln Z) + \sum_id\lambda_i\langle A_i \rangle + \sum_i\lambda_id\langle A_i\rangle ~, 
\end{equation}
then subsequently
\begin{equation}\label{eq:differential of S}
	dS = k \sum_i\lambda_id\langle A_i \rangle ~.
\end{equation}
It can be seen that Eqs.~\eqref{eq:differential of lnZ} and \eqref{eq:differential of S} capture the natural variables for $S$ and $\ln Z$. Since $S$ is a function of $\langle A_i \rangle$, and $\ln Z$ is also useful as a function of $\lambda_i$, one can write
\begin{equation}
	\lambda_i = \frac{1}{k}\frac{\partial S}{\partial \langle A_i\rangle} ~,
\end{equation}
and
\begin{equation}
	\ln Z  = \frac{S}{k} - \frac{1}{k}\sum_i\langle A_i \rangle \frac{\partial  S}{\partial\langle A_i \rangle } ~.
\end{equation}
Therefore, $\lambda_i $ is the conjugate variable of $\langle A_i \rangle$ concerning $S/k$, and $\langle A_i \rangle $ is the conjugate variable of $\lambda_i$ concerning $-\ln Z$.

\section{Derivation of Eqs. (14) and (15)}\label{appC}
In the context of quantum-enhanced parameter estimation, a suitable measure to determine the ultimate limit of estimation precision is the Cram\'er-Rao (CR) bound \cite{fanchini2017lectures,cavina2018bridging,liu2019quantum}
\begin{equation}
	\Delta{\boldsymbol{\theta}}\geq{\frac{1}{\sqrt{n\mathcal{F}(\theta)}}}~,
\end{equation}
with $\mathcal{F}(\theta)$ being the QFI and $\theta$ standing for the unknown parameter encoded in the initial state. By considering $\rho = e^G$ and $G=-\sum_i\lambda_iA_i - \ln Z$, where $\lambda_i$ are the Lagrange coefficients, and $A_i\,|e_j\rangle = a_{ij}\,|e_j\rangle$, we can write the QFI in the following form \cite{modi2011quantum,hayashi2016quantum,fiderer2019maximal}
\begin{equation}
	\mathcal F(\lambda_p)=2\sum_{n,m}\frac{|\langle e_n|\partial_{\lambda_p}\rho|e_m\rangle|^2}{p_n+p_m}~,
\end{equation}
for the desired Lagrange coefficient $\lambda_p$. Regarding Wilcox's formula for the derivative of an exponential operator \cite{jiang2014quantum}
\begin{equation}
	\partial_{\lambda_{p}}e^{G} = \int_0^1 e^{\alpha G}\,\partial_{\lambda_{p}}G \,e^{(1-\alpha)G}\,d\alpha ~,
\end{equation}
and spectral decomposition $\rho=\sum_n p_n |e_n\rangle\langle e_n|$, the QFI becomes 
\begin{align}\label{SSS}
	\begin{split}
		\mathcal F(\lambda_p)&=2\sum_{n,m}\frac{1}{p_n+p_m}\left|\langle e_n|\int_0^1e^{\alpha G}\partial_{\lambda_p}[G] e^{(1-\alpha )G}|e_m\rangle\right|^2\\
		&=2\sum_{n,m}\frac{1}{p_n+p_m}|\langle e_n|\partial_{\lambda_p} [G]|e_m\rangle|^2\left|\int_0^1e^{\alpha g_n+(1-\alpha)g_m}\right|^2\\
		&=\sum_{n} \frac{1}{p_n}|\langle e_n|\partial_{\lambda_p}[G]|e_n\rangle|^2 e^{2g_n}\\ &+2\sum_{n\neq m}\frac{1}{p_n+p_m}|\langle e_n|\partial_{\lambda_p} [G]|e_m\rangle|^2\left(\frac{e^{g_m}-e^{g_n}}{g_m-g_n}\right)^2 ~.
	\end{split}
\end{align}
Here, we have used the fact that $G\,|e_j\rangle = g_j\,|e_j\rangle$, where $g_j = - \lambda_{j} a_{j} - \ln Z$. Also, it is substantial to keep the following in mind
\begin{equation}\label{SS1}
	\begin{split}
		\frac{\partial G}{\partial \lambda_p}&=\frac{\partial}{\partial\lambda_p}\left(-\sum_l \lambda_l A_l -\ln Z\right)\\
		&=\left(-\sum_l\frac{\partial \lambda_l}{\partial \lambda_p}A_l - \frac{\partial \ln Z}{\partial \lambda_p}\right)\\
		&=-A_p +\langle A_p\rangle=-\delta A_p ~.
	\end{split}
\end{equation}
In the second line of Eq.~\eqref{SS1}, we assumed that ${\lambda_l}$'s with $l\neq p$ and $A_p$ are independent of $\lambda_p$. Clearly, this assumption cannot be used for the strong coupling regime in which the energy operator has a dependence on temperature (its own Lagrange coefficient).
By rewriting the GGE state in Eq.~\eqref{S} as
\begin{align}
	\rho =\sum_ne^{G}|e_n\rangle\langle e_n|&= \sum_n e^{-\sum_i\lambda_iA_i-\ln Z}|e_n\rangle\langle e_n|\nonumber \\&=\sum_n e^{-\sum_i\lambda_ia_{ni}-\ln Z}|e_n\rangle\langle e_n|\nonumber\\&=\sum_ne^{g_n}|e_n\rangle\langle e_n|=\sum_n p_n|e_n\rangle\langle e_n| ~,
\end{align}
and substituting $g_n=\ln(p_n)$ together with Eq.~\eqref{SS1} and by regarding $K_\alpha(\rho,G)= \mathrm{Var}[\rho,G]=\sum_n\langle e_n| e^G (\delta G)^2 |e_n\rangle$, it turns out that
\begin{align}
	\mathcal F(\lambda_p)=\sum_n p_n|\langle e_n|\delta A|e_n\rangle|^2+2\sum_{n\neq m}\frac{(p_m-p_n)^2}{(p_n+p_m)(\ln(p_m/p_n))^2}|\langle e_n|\delta A_p |e_m\rangle|^2 ,
\end{align}
where $$\langle e_n|\delta A_p |e_m\rangle = a_{pm}\delta_{nm}-\langle A_p \rangle \delta_{nm}~,$$
therefore, one comes to
\begin{equation}
	\mathcal F(\lambda_p)=\sum_np_n( a_{pn}^2-2a_{pn}\langle A_p\rangle + \langle A_p\rangle^2) = \sum_n p_n (a_{pn}-\langle A_p\rangle )^2 ~.
\end{equation}
At this point, let us consider the definition of classical uncertainty contribution
\begin{align}
	K_\alpha(\rho,G)= \mathrm{Var}(\rho,G)=\langle (\delta G)^2\rangle = \sum_n\langle e_n| e^G (\delta G)^2 |e_n\rangle ~.
\end{align}
Var() can be written as
\begin{align}
	\mathrm{Var}(\rho,G) &= \sum_{m,n,p}\langle e_n|e^{g_p}|e_p\rangle\langle e_p| \delta G |e_m\rangle\langle e_m|\delta G |e_n \rangle \nonumber\\
	&=\sum_{m,n}p_n \langle e_n|\delta G \vert e_m\rangle\langle e_m|\delta G|e_n\rangle \nonumber\\
	&=\sum_n p_n |\langle e_n|\delta G|e_n\rangle|^2 + \sum_{n\neq m}p_n |\langle e_n|\delta G|e_m\rangle|^2 ~,
\end{align}
where $\delta G= -\sum_i \lambda_i \delta A_i$, hence
\begin{align}
	\mathrm{Var}(\rho,G) &= \sum_{n,i}p_n |\langle e_n | \lambda_i \delta A_i|e_n\rangle|^2 +\sum_{n\neq m,i}p_n |\langle e_n | \lambda_i \delta A_i|e_n\rangle|^2\nonumber\\
	&=\sum_{n,i}p_n\lambda_i^2(a^2_{ni}-2a_{pn}\langle A_i\rangle +\langle A_i\rangle^2)\nonumber\\
	&=\sum_{n,i}p_n \lambda_i^2(a_{ni}-\langle A_i \rangle )^2 ~.
\end{align}
\section{Derivation of Eq. (22)}\label{appD}
According to the previous methods
\begin{align}
	\begin{split}
		\mathcal F(\lambda_p)&=\sum_{n,m}\frac{2}{p_n+p_m}\left|\langle e_n|\int_0^1e^{\alpha G^*_S}\partial_{\lambda_p}G^*_S e^{(1-\alpha)G^*_S}|e_m\rangle\right|^2\\
		&=\sum_{n,m,l,q}\frac{2}{p_n+p_m}\left|\langle e_n|\int_0^1e^{\alpha g_l}|e_l\rangle\langle e_l|\partial_{\lambda_p}G^*_S e^{(1-\alpha)g_q}|e_q\rangle\langle e_q|e_m\rangle\right|^2\\
		&=2\sum_{n,m}\frac{1}{p_n+p_m}|\langle e_n|\partial_{\lambda_p}G^*_S |e_m\rangle|^2\left|\int_0^1e^{\alpha g_n+(1-\alpha)g_m}\right|^2\\
		&=\sum_{n} \frac{1}{p_n}|\langle e_n|\partial_{\lambda_p}G^*_S |e_n\rangle|^2 e^{2g_n} +2\sum_{n\neq m}\frac{1}{p_n+p_m}|\langle e_n|\partial_{\lambda_p}G^*_S |e_m\rangle|^2\left(\frac{e^{g_m}-e^{g_n}}{g_m-g_n}\right)^2 ~,
	\end{split}
\end{align}
where we have used the fact that $G^*_S|e_j\rangle = g_j|e_j\rangle $ and
\begin{equation}\label{56}
	\begin{split}
		\frac{\partial G^*_S}{\partial \lambda_p}&=\frac{\partial}{\partial\lambda_p}\left(-\sum_i \lambda_i A^*_i -\ln Z^*_S\right)\\
		&=-\frac{\partial}{\partial\lambda_p}\sum_i\lambda_iA^*_i +\left\langle\frac{\partial}{\partial\lambda_p}\left(\sum_i\lambda_iA^*_i\right)\right\rangle\\
		&=-E^*_S +\langle E^*_S\rangle=-\delta E^*_S ~.
	\end{split}
\end{equation}
It is also shown that
\begin{align}\label{57}
	\rho_S &=\sum_ne^{G^*_S}|e_n\rangle\langle e_n| = e^{-\sum_i\lambda_iA^*_i-\ln Z^*_S}|e_n\rangle\langle e_n|\nonumber \\ &=\sum_n e^{-\sum_i\lambda_ia_{ni}-\ln Z^*_S}|e_n\rangle\langle e_n|=\sum_ne^{g_n}|e_n\rangle\langle e_n|\nonumber\\&=\sum_np_n|e_n\rangle\langle e_n|~.
\end{align}
Implementing Eqs.~\eqref{56} and \eqref{57} as well as $g_n=\ln(p_n)$, leads to
\begin{align}
	\mathcal F(\lambda_p)=\sum_n p_n|\langle e_n|\delta E^*_S|e_n\rangle|^2+2\sum_{n\neq m}\frac{(p_m-p_n)^2}{(p_n+p_m)(\ln(p_m/p_n))^2}|\langle e_n|\delta E^*_S |e_m\rangle|^2 ~,
\end{align}
and by applying $|\langle e_n|\delta E^*_S |e_m\rangle| = |\langle e_n| E^*_S |e_m\rangle|$ for $n\neq m$, we have
\begin{align}
	\mathcal F(\lambda_p)=\sum_n p_n|\langle e_n|\delta E^*_S|e_n\rangle|^2+2\sum_{n\neq m}\frac{(p_m-p_n)^2}{(p_n+p_m)(\ln(p_m/p_n))^2}|\langle e_n| E^*_S |e_m\rangle|^2 ~.
\end{align}
Given that 
\begin{equation}
	\mathrm{Var}(\rho_S,E^*_S)=\langle (\delta E^*_S)^2\rangle = \sum_n\langle e_n|e^{G^*_S}(\delta E^*_S)^2|e_n\rangle ~,
\end{equation}
one obtains
\begin{align}
	\begin{split}
		\mathrm{Var}(\rho_S,E^*_S)&=\frac{1}{Z_S^*(\{\lambda_i\})}\sum_{m,n,p}\langle e_n|e^{-\sum_i\lambda_ia_{pi}}|e_p\rangle\langle e_p|\delta E^*_S|e_m\rangle\langle e_m|\delta E^*_S |e_n\rangle\\
		&=\sum_{m,n}p_n\langle e_n|\delta E^*_S|e_m\rangle\langle e_m| \delta E^*_S |e_n\rangle\\
		&=\sum_n p_n |\langle e_n|\delta E^*_S|e_n\rangle|^2+\sum_{m\neq n}p_n|\langle e_n| E^*_S|e_m\rangle|^2 .
	\end{split}
\end{align}
Subsequently, the QFI is given as the following form
\begin{align}
	\mathcal F(\lambda_p)=\mathrm{Var}(\rho_S,E^*_S)+\sum_{n\neq m}\left[\frac{2(p_m-p_n)^2}{(p_n+p_m)(\ln(p_m/p_n))^2}-p_n\right]|\langle e_n| E^*_S |e_m\rangle|^2 ~.
\end{align}

\section{Derivation of Eq. (24)}\label{appE}
On the one hand, the skew information can be derived in the following way
\begin{align}
	Q(\rho_S,E^*_S) = \int_0^1 d\alpha Q_\alpha(\rho_S,E^*_S)=-\frac{1}{2}\int_0^1 d\alpha  \mathrm{tr}\left\{[E^*_S,\rho_S^\alpha][E^*_S,\rho_S^{1-\alpha}]\right\} ~,
\end{align}
where
\begin{align}
	\begin{split}
		\mathrm{tr}\left\{[{E^*_S},\rho_S^\alpha][E^*_S,\rho_S^{1-\alpha}]\right\}&=\mathrm{tr}\left\{[{E^*_S}\rho_S^\alpha-\rho_S^\alpha E^*_S][E^*_S\rho_{S}^{(1-\alpha)}-\rho_{S}^{(1-\alpha)}E^*_S]\right\}\\
		&=\mathrm{tr}\left\{E^*_S\rho_S^{\alpha}E^*_S\rho_S^{1-\alpha}\right\}-\mathrm{tr}\left\{E^*_S\rho_S^{\alpha}\rho_S^{1-\alpha}E^*_S\right\}-\mathrm{tr}\left\{\rho_S^{\alpha}E^*_SE^*_S\rho_S^{1-\alpha}\right\}\nonumber\\&+\mathrm{tr}\left\{\rho_S^{\alpha}E^*_S\rho_S^{1-\alpha}E^*_S\right\}~.
	\end{split}
\end{align}
On the other hand, knowing that $\rho_S^\alpha \rho_S^{1-\alpha}=\rho_S$ and that a trace has the cyclic permutation property, we find
\begin{align}
	\mathrm{tr}(E^*_S\rho_S^\alpha \rho_S^{1-\alpha}E^*_S)=\mathrm{tr}({E^{*}_{S}}^{2}\rho_S) = \langle {E^*_S}^{2} \rangle ~.
\end{align}
Also, by utilizing this fact
\begin{align}
	\mathrm{tr}(E^*_S\rho^{\alpha}_S E^*_S\rho^{1-\alpha}_S)=\mathrm{tr}(\rho_S^\alpha E^*_S\rho_S^{1-\alpha}E^*_S)~,
\end{align}
we obtain
\begin{align}
	\begin{split}
		\int_0^1 d\alpha\,\mathrm{tr}(E^*_S\rho_S^\alpha E^*_S\rho_S^{1-\alpha})&=\sum_n\langle e_n|E^*_S \int_0^1 d\alpha\,e^{-G^*_S}E^*_S e^{-(1-\alpha)G^*_S}|e_n\rangle\\
		&=\sum_{n,m,l}\langle e_n|E^*_S\int_0^1dae^{-\alpha g_m}|e_m\rangle\langle e_m|E^*_S e^{-(1-\alpha)g_l}|e_l\rangle\langle e_l|e_n\rangle \\
		&=\sum_{n,m}\langle e_n|E^*_S|e_m\rangle\langle e_m|E^*_S|e_n\rangle \int_0^1 d\alpha\,e^{[\alpha g_m+(1-\alpha)g_n]}\\
		&=\sum_n e^{-g_n}|\langle e_n|E^*_S|e_n\rangle|^2 + \sum_{n\neq m} |\langle e_n|E^*_S|e_m\rangle|^2\left(\frac{e^{g_m}-e^{g_n}}{g_m-g_n}\right)\\
		&=\sum_n p_n|\langle e_n|E^*_S|e_n\rangle|^2+\sum_{n\neq m}\frac{p_m-p_n}{\ln(p_m/p_n)} |\langle e_n|E^*_S|e_m\rangle|^2 ~,
	\end{split}
\end{align}
hence 
\begin{align}
	Q(\rho_S,E^*_S)= -\frac{1}{2}\left[2\sum_n p_n|\langle e_n|E^*_S|e_n\rangle|^2\right]-\frac{1}{2}\left[2\sum_{n\neq m}\frac{p_m-p_n}{\ln(p_m/p_n)} \langle e_n|E^*_S|e_m\rangle|^2 - 2\langle {E^{*}_{S}}^2\rangle \right]~.
\end{align}
One can translate $\langle {E^{*}_{S}}^2\rangle$ by
\begin{align}
	\langle {E^{*}_{S}}^2\rangle &= \sum_{n,m,l} \langle e_l|e^{g_n}|e_n\rangle\langle e_n |E^*_S|e_m\rangle \langle e_m |E^*_S|e_n\rangle \nonumber\\
	&=\sum_{n,m}p_n|\langle e_n |E^*_S|e_m\rangle|^2\nonumber\\
	&=\sum_{n}p_n|\langle e_n|E^*_S|e_n\rangle|^2 +\sum_{n\neq m}p_n|\langle e_n|E^*_S|e_m\rangle|^2 ~.
\end{align}
Therefore
\begin{align}
	Q(\rho_S,E^*_S)=\sum_{n\neq m}\left(p_n-\frac{p_m-p_n}{\ln(p_m/p_n)}\right)|\langle e_n|E^*_S|e_m\rangle|^2 ~.
\end{align}

\section{Derivation of Eqs. (25) and (31)}\label{appF}
By adding the terms related to the skew information and the QFI, we achieve
\begin{align}\label{72}
	\mathcal F(\lambda_p)+Q(\rho_S,E^*_S)=\mathrm{Var}(\rho_S,E^*_S)+\sum_{n\neq m}\left[\frac{2(p_m-p_n)^2}{(p_n+p_m)(\ln(p_m/p_n))^2}-\frac{p_m-p_n}{\ln(p_m/p_n)}\right]|\langle e_n|E^*_S|e_m\rangle|^2 ~.	
\end{align}
Now, the following factorization can be considered
\begin{align}\label{ex}
	\left[\frac{2(p_m-p_n)}{(p_n+p_m)(\ln(p_m/p_n))}-1\right]\left(\frac{p_m-p_n}{\ln(p_m/p_n)}\right)~,
\end{align}
where the second term in the multiplication is always positive, since  $p_m,p_n\in [0,1]$. Thus, the condition for the expression \eqref{ex} to be positive is assigned to the following inequality
\begin{align}
	\frac{(p_m-p_n)}{(p_n+p_m)}&>\frac{1}{2}\ln(p_m/p_n)~.
\end{align}
One can follow the solution in two different situations: The first one is when
\begin{align}\label{I}
	\frac{x-1}{x+1}>\frac{1}{2}\ln x, \qquad x=\frac{p_m}{p_n}~.
\end{align}
The inequality in Eq.~\eqref{I} is satisfied for the interval $0<x<1$, where $p_m<p_n$. 

The second one is when $p_m>p_n$, and we have
\begin{align}\label{Ineq.Case2}
	\frac{1-x}{1+x}>-\frac{1}{2}\ln(x)~,
\end{align}
with $x=p_n/p_m$, which is exactly the same inequality as before with only the sign changed. Therefore, the positivity of the second term on the right side of Eq.~\eqref{72} boils down to the inequalities \eqref{I} and \eqref{Ineq.Case2}, which were proven to be valid for any real values of $p_m$ and $p_n$ between 0 and 1.
From these considerations, we can write an inequality as
\begin{align}\label{F6}
	\mathcal F(\lambda_p) \leq \mathrm{Var}(\rho_S,E^*_S)-Q(\rho_S,E^*_S).
\end{align}
Furthermore, we express the uncertainty using the operators $A^*_i$'s by
\begin{align}
	E^{*p}_{S}=\sum_i\frac{\partial(\lambda_iA^*_i)}{\partial\lambda_p}~,
\end{align}
then
\begin{align}
	K(\rho_S,E^{*p}_S)&=\int_0^1d\alpha\,\mathrm{tr}\{\rho_S^\alpha \delta E^{*p}_S\rho_S^{1-\alpha}\delta E^{*p}_S\}~,
\end{align} 
where $\delta E^{*p}_S=E^{*p}_S-\langle E^{*p}_S\rangle$. By considering the dependence $\lambda_i$ and $A^*_p$ on $\lambda_p$, we have
\begin{align}
	E^{*p}_{S} = \sum_{i\neq p}(\dot{\lambda}_iA_i^*+\lambda_i\dot{\lambda}_i\frac{\partial A_i^*}{\partial \lambda_i})+\partial_{\lambda_p}[\lambda_p A_p^*], \qquad \mathrm{with} \quad \dot{\lambda}_i=\frac{\partial \lambda_i}{\partial \lambda_p}~.
\end{align}
Finally, given that $[A^*_p,\rho_S]=0~(i\neq p)$, we derive a general expression for classical uncertainty in the following form
\begin{align}
	K(\rho_S, E^{*p}_{S}) &= \int_0^1d\alpha \mathrm{tr}\left\{\rho_S^\alpha[\partial_{\lambda_p}[ {\lambda_p}A^*_p]\rho_S^{1-\alpha}\partial_{\lambda_p} [{\lambda_p}A^*_p]\right\}- \langle\partial_{\lambda_p}[ \lambda_p A^*_p]\rangle^2+\sum_{i\neq p}\dot{\lambda}_i\langle\partial_{\lambda_p}[ {\lambda_p}A^*_p][A^*_i+\lambda_i \frac{\partial A^*_i}{\partial\lambda_i}]\rangle\nonumber\\&+\sum_{j\neq p}\dot{\lambda}_j\langle \partial_{\lambda_p}[\lambda_p A^*_p][A^*_j+\lambda_j\frac{\partial A^*_j}{\partial\lambda_j}]\rangle-\sum_{i\neq p}\dot{\lambda}_i\langle \partial_{\lambda_p} [{\lambda_p}A^*_p]\rangle~\langle A^*_i+\lambda_i \frac{\partial A^*_i}{\partial\lambda_i}\rangle \nonumber\\
	&-\sum_{j\neq p}\dot{\lambda}_j\langle\partial_{\lambda_p}[ {\lambda_p}A^*_p]\rangle~\langle A^*_j+\lambda_j\frac{\partial A^*_j}{\partial\lambda_j}\rangle+\sum_{ij\neq p}\dot{\lambda}_i\dot{\lambda}_j \langle [A^*_i+\lambda_i \frac{\partial A^*_i}{\partial\lambda_i}][A^*_j+\lambda_j\frac{\partial A^*_j}{\partial\lambda_j}] \rangle\nonumber\\
	&-\sum_{ij\neq p}\dot{\lambda}_i\dot{\lambda}_j\langle A^*_i+\lambda_i \frac{\partial A^*_i}{\partial\lambda_i}\rangle~ \langle A^*_j+\lambda_j\frac{\partial A^*_j}{\partial\lambda_j}\rangle\nonumber\\
	&=K(\rho_S,\partial_{\lambda_p}[\lambda_p A^*_p])+2\sum_{i\neq p} \dot{\lambda}_i~ \mathrm{Cov}~(\partial_{\lambda_p}[\lambda_p A^*_p],~A^*_i+\lambda_i \frac{\partial A^*_i}{\partial\lambda_i})\nonumber\\&+\sum_{ij\neq p}\dot{\lambda}_i\dot{\lambda}_j~ \mathrm{Cov}~(A^*_i+\lambda_i \frac{\partial A^*_i}{\partial\lambda_i},~A^*_j+\lambda_j\frac{\partial A^*_j}{\partial\lambda_j})~.
\end{align}
\textcolor{red}{where $\mathrm{Cov}(X,Y)=\langle X Y\rangle-\langle X\rangle \langle Y\rangle$.}

\end{document}